\documentclass[tighten,flushrt,preprint]{aastex}  

\def\eq#1{equation~(\ref{#1})}

\def\ut{{\tilde u}}

\font\BFd=cmmib10 scaled 1200
\font\BFt=cmmib10 scaled 1200
\font\BFs=cmmib10 

\def\bb#1{\relax 
\ifmmode\mathchoice
{{\hbox{\BFd #1}}}{{\hbox{\BFt #1}}}
{{\hbox{\BFs #1}}}{{\hbox{\BFs #1}}}
\else \mbox{#1} \fi }

\def\x{{\bb{x}}}
\def\y{{\bb{y}}}
\def\k{{\bb{k}}}
\def\rhobar{{\bar \rho}}
\def\deltabar{{\bar \delta}}
\def\xibar{{\bar \xi}}

\begin{document}

\title{What Does It Take to Stabilize Gravitational Clustering?}

\author{Chung-Pei Ma}

\affil{Department of Physics and Astronomy, University of Pennsylvania, 
Philadelphia PA 19104; cpma@physics.upenn.edu}

\and

\author{J. N. Fry}

\affil{Department of Physics, University of Florida, 
Gainesville FL 32611-8440; fry@phys.ufl.edu}

\begin{abstract}

An analytical understanding of the strongly nonlinear regime of
gravitational collapse has been difficult to achieve.  The only
insight has been the stable clustering hypothesis, which assumes that
the number of neighbors for objects averaged over small length scales
is constant in time.  Our recently proposed analytic halo model for
$N$-point correlation functions now provides a tool for calculating
gravitational clustering properties in the strongly nonlinear regime.
This model also provides a new physical framework for an independent
evaluation of the validity of the stable clustering hypothesis.  We
derive the asymptotic nonlinear behavior of the $N$-point correlation
functions and pairwise peculiar velocities in terms of dark matter
halo properties.  We show that these statistics exhibit stable
clustering only when the halo mass function and halo density profile
obey specific relations.  The long-cherished stable clustering
hypothesis therefore is not necessarily realized in practice.

\end{abstract}

\subjectheadings{cosmology : theory -- dark matter -- large-scale
structure of universe}

\section{Introduction}

Gravitational clustering is the fundamental physical process
responsible for the formation and evolution of structure in the
universe.  Galaxy-hosting dark matter halos are the products of the
strongly nonlinear stage of this process, but a detailed understanding
of this important regime has mostly eluded us because few numerical
simulations have the dynamic range to explore comfortably such a small
length scale (e.g., Moore et al. 1999; Bullock et al. 1999).  The one
analytic handle has been the stable clustering assumption (Peebles
1974; Davis \& Peebles 1977; Peebles 1980), which has allowed us to
extrapolate into nonlinear regimes beyond the reach of numerical
simulations.  Predictions of the stable clustering hypothesis are: (1)
the two-point correlation function of the density field $\xi(r)$ and
the power spectrum $P(k)$ are power laws, with $\xi(r) \propto
r^{-\gamma}$ and $\Delta(k)\equiv 4\pi k^3 P(k) \propto k^\gamma$,
where $\gamma = {(9+3n)/(5+n)}$ for a primordial spectral index $n$;
(2) the higher order $N$-point correlation functions $\xi_N(r)$ scale
as $\xi_N(r) \propto r^{-\gamma_N}$, with $\gamma_N = (N-1)\gamma$, or
$ \xi_N \propto \xi^{N-1} $; and (3) the pairwise peculiar velocity
exactly cancels the Hubble expansion on small scales, $v/Hr=-1$, so
that bound, high-density halos maintain a fixed physical size.  Some
of these predictions have agreed well enough with numerical simulation
results (Jain 1997 and references therein) to have obtained general
acceptance.

In two recent papers we presented evidence from high-resolution
$N$-body simulations that gravitational clustering does not
necessarily follow the scaling $\xi_3\propto \xi^2$ required by the
stability condition for the two- and three-point functions $\xi$ and
$\xi_3$ (Ma \& Fry 2000a).  For a deeper understanding of the strongly
nonlinear regime beyond simulations, we proceeded to construct an
analytic halo model (Ma \& Fry 2000b) in which mass is distributed in
spherical dark matter halos with phenomenological mass distribution
functions, density profiles, and halo-halo correlations.  We showed
that this halo model can reproduce analytically the two- and
three-point correlation functions measured in numerical simulations,
and that it also makes useful predictions beyond the limited range of
validity of simulations.

In this {\it Letter} we investigate the extent to which the analytic
halo model is consistent with stable gravitational clustering in the
strongly nonlinear regime.  In particular, we derive the halo model
predictions for the asymptotic nonlinear behavior of the $N$-point
correlation functions (\S 2) and the pairwise peculiar velocities (\S
3) in terms of dark matter halo properties.  We then obtain the
conditions that must be satisfied by the halo mass function and
density profile in order to reproduce results of the stable clustering
hypothesis.

\section{$N$-Point Correlation Functions}

The analytic halo model of Ma \& Fry (2000b) for the $N$-point
correlation functions of the mass density takes three inputs: the halo
mass distribution function $dn/dM$; the halo density profile
$\rho(r)/\rhobar=\deltabar\,u(r/R_s)$ where $R_s$ is a scale radius
and $\deltabar$ is the normalization; and halo-halo correlations.  In
this model, the two-point function contains contributions from
particle pairs residing in a single halo and in two separate halos,
$\xi(r) = \xi_{1h}(r) + \xi_{2h}(r)\,$, where the subscripts ``$1h$''
and ``$2h$'' denote the respective ``1-halo'' and ``2-halo''
contributions (see also Seljak 2000).  Similarly, we write the
bispectrum $B(k_1,k_2,k_3)$ as three separate terms, representing
contributions from particle triplets that reside in a single halo and
in two and three distinct halos.  As shown in Ma \& Fry (2000b), the
dominant contribution in the nonlinear regime on small length scales
is from the ``1-halo'' terms for particles that reside in the same
halos.  This makes intuitive sense because closely spaced particles
are most likely to be found in the same halo.  These ``1-halo'' terms
for the two- and $N$-point statistics are
\begin{eqnarray}
  &&\xi_{1h}(r) = \int dM\,{dn\over dM}\,
        R_s^3\,\deltabar^2\, \lambda_2(r/R_s) \,, \qquad
   P_{1h}(k) = \int dM\,{dn\over dM} \,
        [ R_s^3\,\deltabar\,\ut (kR_s)]^2  \,, \nonumber \\
 &&\xi_{N,1h}(r_1,\cdots,r_N) = \int dM\,{dn\over dM}\,
        R_s^3\,\deltabar^N\, \lambda_N(r_1/R_s,\cdots,r_N/R_s) \,, 
   \label{1-halo} \\
  && P_{N,1h}(\k_1, \dots, \k_N) = \int dM\,{dn\over dM}
    \, [R_s^3\,\deltabar\,\ut(k_1 R_s)] \, \cdots 
    \, [R_s^3\,\deltabar\, \ut(k_N R_s)] 
    \,. \nonumber
\end{eqnarray}
where 
$\lambda_N(x_1,\cdots,x_N)\equiv \int d^3y\, u(y)\,
u(|\y+\x_1-\x_2|)\cdots u(|\y+\x_1-\x_N|)$; $\ut$ is the Fourier
transform of $u(x)$; the $N$-point $P_N$ is defined for
$\sum \k_i = 0$, and $P_3=B$ is the bispectrum.

For our goal of studying the $N$-point functions in the strongly
nonlinear regime, we need to consider only the halo mass function and
halo density profile and not halo-halo correlations because the latter
make negligible contributions on small length scales and are therefore
irrelevant for stable clustering (see Ma \& Fry 2000b for the full
expressions for $P$ and $B$).  For the halo mass function, we write it
as
\begin{equation}
  {dn\over dM} \propto M^{-2}\,\nu^\alpha\, e^{-\nu^2/2}\,,\qquad 
	\nu={\delta_c\over \sigma(M)} \,,
\label{alpha}
\end{equation}
where $\alpha$ parameterizes the uncertainty in the logarithmic slope
at the low-mass end, $\delta_c\approx 1.68$ characterizes the linear
overdensity at the onset of gravitational collapse, and $\sigma(M)$ is
the linear rms mass fluctuations in spheres of radius $R$, where
$M=4\pi\bar\rho R^3/3$.  Values in use for $\alpha$ vary from 1 (Press
\& Schechter 1974) to about 0.4 (Sheth \& Tormen 1999; Jenkins et
al. 2000).  For the halo density profile, we assume a ``universal''
shape, $\rho/\bar{\rho}=\deltabar\,u(r/R_s)$, where $u=1/[x(1+x)^2]$
is proposed by Navarro et al. (1997), while $u(x)=1/(x^{3/2}+x^3)$ is
suggested by Moore et al. (1999).  Both the scale radius $R_s$ and the
normalization $\deltabar$ can be expressed in terms of a concentration
parameter $c(M)\equiv R_{200}/R_s$, where $R_{200}$ is the virial
radius within which the average halo density is 200 times the mean
density of the universe.  Specifically, $R_s\propto M^{1/3}/c$, and
$\deltabar \propto c^3\,g(c)$, where $g(c)$ is a logarithmic factor
with $g=1/[\ln(1+c)-c/(1+c)]$ for Navarro et al. (1997) and
$g=1/\ln(1+c^{3/2})$ for Moore et al. (1999).  High resolution
simulations of individual halos show a roughly power-law $c(M)$ with
some, perhaps substantial, scatter (e.g., Cole \& Lacey 1996; Tormen
et al 1997; Navarro et al. 1997; Jing \& Suto 2000), so we will write
\begin{equation}
  c \equiv {R_{200} \over R_s } \propto \left( {M\over M^*} 
  \right)^{-\beta_0} , \label{beta}
\end{equation}
where $M^*$ is defined by $\sigma(M^*)=1$, and $\beta_0$ is used to
parameterize the uncertainty in the mass dependence, which has been
found to be in the range $0\la \beta_0 \la 1/2$.

We now derive analytically the small-$r$ behavior of \eq{1-halo}, in
which the factors $dn/dM$, $R_s$, and $\deltabar$ are all functions of
the halo mass $M$.  For small $r$ or high $k$, the integral for
$\xi_{1h}(r)$ or $P_{1h}(k)$ converges before the exponential cutoff
in the mass function, and the dominant contribution to the integral
comes from masses near the scale for which $r/R_s=1$ or $k R_s=1$.  We
then have $P(k) \approx P_{1h}(k) \propto \int dM\,\nu^\alpha \,
\ut^2(kR_s)\,g^2(c)$ at high $k$.  Changing variables to $y = kR_s
\propto k\,M^{\beta_0 +1/3}$, we obtain the asymptotic behavior
\begin{equation}
   \xi(r)\propto r^{-\gamma}\,,\quad
  \Delta(k)\equiv 4\pi k^3\, P(k) \propto k^\gamma\,,\qquad \gamma = 
	{18\beta - \alpha\,(n+3) \over 2\,(3\beta+1)} \,,
\label{gamma_model}
\end{equation}
where we have $\beta=\beta_0$ if the logarithmic factor $g(c)$ in
$\deltabar$ is ignored, or more accurately, we have $\beta\approx
0.8\,\beta_0$ if the effect of $g(c)$ is approximated by a power law.
From \eq{gamma_model}, we see that in order to reproduce the two-point
stable clustering result for a given spectral index $n$, the indices
$\alpha$ and $\beta$ must satisfy
\begin{equation}
  {18\beta - \alpha\,(n+3) \over 2\,(3\beta+1)} = {9+3n\over 5+n} \,.
  \label{stable2}
\end{equation}
For $n=-2$, for example, this is satisfied if $\beta=0.25$ or
$\beta_0\approx 0.31$ for $\alpha=1$, and $\beta=0.2$ or
$\beta_0\approx 0.25$ for $\alpha=0.4$.

For the three-point function $\xi_3$ and the bispectrum $B$, the
integrals in \eq{1-halo} converge if $\epsilon <3/2$ for an
inner halo density of $r^{-\epsilon}$.  For $\xi_3$, $B$, and the
commonly defined three-point amplitude $Q\propto B/P^2$, we find
\begin{equation}
	\xi_3(r) \propto r^{-\gamma_3} \,, \quad
     k^6 B(k) \propto  k^{\gamma_3} \,,\quad 
     Q(k) \propto  k^{\gamma_3-2\gamma}\,,\quad
    \gamma_3 = {36\beta - \alpha\,(n+3) \over 2\,(3\beta+1)} \,.
\label{gamma3_model}
\end{equation}
Stable clustering requires a scale-independent $Q(k)$, and 
\begin{equation}
	\gamma_3-2\gamma= {\alpha\,(n+3) \over 2\,(3\beta+1)} = 0 \,,
\label{stable3}
\end{equation}
which holds only if $\alpha=0$ or $n=-3$.  Combining
equations~(\ref{stable2}) and (\ref{stable3}), we find that stable
clustering requires $\alpha=0$ and $\beta=(3+n)/6$.  All work on halo
mass functions performed thus far has reported a much larger value for
$\alpha$; stable clustering therefore does not appear to be followed,
which is consistent with the scale-dependent $Q(k)$ reported in Ma \&
Fry (2000a).

We illustrate these results by showing in Figure~1 the power spectrum
and bispectrum computed from the analytic halo model for three sets of
($\alpha$, $\beta$) for the $n=-2$ scale-free model.  The values of
$\alpha$ and $\beta$ are all chosen to give the same slope $\gamma=1$
for the nonlinear two-point function; the corresponding power spectra
are therefore nearly identical at high $k$.  The three-point $Q$,
however, have very different high-$k$ behavior for the three cases,
and only the case $\alpha=0$ produces an approximately
scale-independent $Q$.  ($Q$ is not perfectly flat for
$(\alpha,\beta)=(0,1/6)$ because as mentioned above, the simple
expressions in eqs.~(\ref{stable2}) and (\ref{stable3}) are derived by
approximating the logarithmic factor $g(c)$ in the halo profile as a
power law.)

For the four-point and higher correlation functions, provided the
integral is convergent without the exponential cutoff in the mass
function, the same change of variable gives
\begin{equation}
   \xi_N \propto r^{-\gamma_N} \,,\quad 
   k^{3(N-1)}P_N \propto k^{\gamma_N} \,, \quad 
   \gamma_N = {18 (N-1) \beta - \alpha (n+3) \over 2 (3\beta+1)} \,.
\label{gammaN_model}
\end{equation}
Stable clustering requires 
\begin{equation}
	\gamma_N-(N-1)\gamma= (N-2){\alpha\,(n+3) \over 2\,(3\beta+1)} = 0 \,,
\label{stableN}
\end{equation}
which again is satisfied by $\alpha=0$ or $n=-3$.  So if the
three-point correlation follows \eq{stable3}, stable clustering
continues to hold for all orders for which the integral converges. At
large $k$, we are probing the inner cores of halos, where the inner
profile goes as $r^{-\epsilon}$.  The Fourier amplitude $\tilde u$
thus falls off as $\tilde u(k) \propto
(kR_s)^{-(3-\epsilon)}$, and convergence requires $N [1-(3-\epsilon)
(3\beta + 1)/3]-1 + \alpha (n+3)/ 6 < 0$.  For $\epsilon=1$
(Navarro et al.) and $n>-2$, this holds for all $N$, but
for $\epsilon={3 \over 2}$ (Moore et al.) and $n=-2$, it
holds only as far as $N=3$.

\section{Pairwise Peculiar Velocities}

Besides the scaling properties of the $N$-point correlation functions
discussed above, stable clustering also predicts that the mean
peculiar velocity $v$ between pairs of objects of separation $r$ must
cancel the Hubble expansion $Hr$ on small scales so that bound halos
maintain a fixed physical size.  The velocity ratio $v/Hr$ is related
to the two-point correlation function $\xi$ by the pair conservation
equation (Peebles 1980)
\begin{equation}
    -{v\over Hr} = {\xibar \over 3(1+\xi)} \, 
   {\partial\, \ln \xibar \over \partial\,\ln a}\,, 
\label{conserv}
\end{equation}
where $a$ is the expansion factor and $\bar\xi(x)\equiv 3 x^{-3}
\int_0^x\,dx'\, x'^2\,\xi(x')$ is the mean two-point correlation
function interior to $x$.  Stable clustering requires $-v/Hr=1$.

The analytic halo model of Ma \& Fry (2000b) makes specific
predictions for the behavior of $v/Hr$.  As eq.~(\ref{conserv})
indicates, the main task is to compute the time dependence of $\xi$.
Equation~(\ref{1-halo}) shows that the time dependence of $\xi$ enters
through the rms density fluctuation $\sigma$ in $dn/dM$ of
eq.~(\ref{alpha}), and the halo concentration parameter $c(M)$ of
eq.~(\ref{beta}).  We obtain
\begin{equation}
  {\partial\, \xi_{1h} \over \partial\,\ln a} = \int dM\,
      \left[ f(\nu^2-\alpha) + (2 \deltabar'-3 + \lambda')\,c'\,\right] 
	{dn\over dM}\,R_s^3\,\deltabar^2\,\lambda(r/R_s) \,,
\label{dxida}
\end{equation}
where $f\equiv d\ln\sigma/d\ln a\approx\Omega^{0.6}$ is the standard
growth rate of the density field, $\deltabar' \equiv
d\ln\deltabar/d\ln\,c$, $\lambda' \equiv d\ln\lambda/d\ln\,x$, and $c'
\equiv d\ln c/d\ln a$.  Specifically, for the Navarro et al. profile,
$\deltabar(c)= (200\,c^3/3) /[\ln(1+c]-c/(1+c)]$ and $\lambda(x)=8\pi
[(x^2+2x+2)\,\ln(1+x)/(x^2+2x)-1]/(x^3+2x^2)$; and similar expressions
for the Moore et al. profile can be found in Ma \& Fry (2000b).  The
term $c'$ for the time dependence of the concentration parameter is
somewhat uncertain.  Numerical simulation results of Bullock et
al. (1999) give a stronger redshift evolution, $c\propto a$, than
Navarro et al. (1997), while more complicated forms have also been
proposed (e.g., Cooray et al. 2000).  We note that in equation
(\ref{beta}), $c(M)$ has an implicit dependence on $a$ through $M^*$
which roughly agrees with Bullock et al. when $\beta=(3+n)/6$.

The prediction of the halo model for $v/Hr$ can be calculated from
equations~(\ref{1-halo}) and (\ref{dxida}).  Before doing so, it is
useful to derive first the asymptotic behavior of $v/Hr$ in the deeply
nonlinear regime at small $r$.  On such small scales, $\xi\gg1$, and
\eq{gamma_model} gives a power law $\xi\propto r^{-\gamma}$ and
$\xibar=3\,\xi/(3-\gamma)$.  For a scale-free model, the self
similarity solution gives $\partial\xibar/\partial\ln a=-2/(3+n)
\,\partial\xibar/\partial\ln r$.  We find that \eq{conserv} then
reaches
\begin{equation}
   -{v\over Hr} \stackrel{r\rightarrow 0}\longrightarrow 
	{2\over n+3} {18\beta -\alpha\,(n+3) \over 6 + \alpha\,(n+3)}\,.
\label{h_model}
\end{equation}
It can be easily checked that $-v/Hr$ approaches unity, as it should,
for any pair $(\alpha,\beta)$ that satisfies the stable clustering
condition of \eq{stable2} for the two-point function.

Figure~2 shows our results for the general behavior of $-v/Hr$
calculated from the analytic halo model.  An $n=-2$ scale-free power
spectrum is used for illustration.  Although $\xi$ is dominated by the
1-halo term $\xi_{1h}$ of \eq{1-halo} at $\xi\ga 1$, which is the
range of interest for this paper, we improve the accuracy at $\xi < 1$
by adding the linear theory $\xi_{l}\propto r^{-(3+n)} \propto r^{-1}$
to $\xi_{1h}$.  As Figure~2 shows, the resulting $v/Hr$ indeed follows
the linear perturbative prediction at $\xi\la 1$, and \eq{h_model} at
$\xi\gg 1$.  We find $-v/Hr \to 1$ for the same pairs of $(\alpha,
\beta)$ shown in Figure~1, which were chosen to illustrate the stable
clustering scaling of the two-point function.  However, $-v/Hr$
reaches a smaller value at $\xi\gg 1$ for the $(\alpha,\beta)$
suggested in the literature, e.g., $\alpha=1$ or 0.4 for the mass
function (Press \& Schechter 1974; Sheth \& Tormen 1997; Jenkins et
al. 2000) and $\beta=1/6$ (Cole \& Lacey 1996; Navarro et al. 1997),
implying that the pairwise peculiar velocities are too small to cancel
exactly the Hubble expansion.  As a consistency check, we have
computed the derivative $\partial\xibar/\partial a$ using both
eq.~(\ref{dxida}) and $[\xibar(a+\Delta a)-\xibar(a)]/{\Delta a}$.
The two methods give identical results, but eq.~(\ref{dxida}) offers a
more physical insight into the various contributing terms.  For
example, the broad peak in $-v/Hr$ for $1\la \xi \la 100$ is the
result of the term in $\nu^2$ in eq.~(\ref{dxida}), which reflects
that the integral on these scales is dominated by mass scales near
$M^*$ where $\sigma(M^*)=1$.

\section{Summary and Discussion}

In this {\it Letter}, we have derived the asymptotic nonlinear
behavior of the $N$-point correlation functions $\xi_N$
(eqs.~[\ref{gamma_model}], [\ref{gamma3_model}], [\ref{gammaN_model}])
and the pairwise peculiar velocities $-v/Hr$ (eq.~[\ref{h_model}]) in
the framework of the recently proposed analytic halo model of Ma \&
Fry (2000b).  We have shown that their small scale behavior is
consistent with the stable clustering hypothesis only if dark matter
halos satisfy certain criteria.  The two halo parameters whose
variation we have explored are the logarithmic slope $\alpha$ of the
halo mass distribution $dn/dM$ at the low mass end in
eq.~(\ref{alpha}) and the slope $\beta_0$ for the mass dependence of
the halo concentration parameter $c(M)$ in eq.~(\ref{beta}) (recall
$\beta\approx 0.8\beta_0$).  These two parameters are highlighted
because results to date from numerical simulations have indicated a
significant uncertainty, with $0.4\la \alpha \le 1$ and $0\la \beta
\la 1/2$ being plausible values.

 From the derived asymptotic nonlinear behavior, we have obtained
analytically the relations that must be satisfied by $\alpha$ and
$\beta$ in order for stable clustering to occur.
Equations~(\ref{stable2}), (\ref{stable3}), and (\ref{stableN})
summarize the results for the $N$-point correlation function $\xi_N$.
These equations and Figure~1 indicate that although the two-point
stable clustering condition alone is satisfied by an infinite set of
$(\alpha,\beta)$ for a given $n$, the three- and higher-point stable
clustering conditions, $\xi_N\propto \xi^{N-1}\,, N\ge 3$, imply
$\alpha$ must be zero.  Achieving stable clustering to all orders
therefore requires stringent conditions: $\alpha=0$ and
$\beta=(3+n)/6$.  For non-scale-free models (e.g., cold dark matter),
$n$ is the index at the high-$k$ end of the power spectrum.

Compared with the hierarchical model of Davis \& Peebles (1977),
results from this paper and Ma \& Fry (2000b) show that the analytic
halo model makes more general and physical predictions for the
behavior of the $N$-point correlation functions and the pairwise
velocities.  Imposing the stability condition $\alpha=0$ and
$\beta=(3+n)/6$ in equations~(\ref{gamma_model}),
(\ref{gamma3_model}), and (\ref{gammaN_model}), we indeed recover
their results of $\xi\propto r^{\gamma}$ with $\gamma=(9+3n)/(5+n)$ at
small $r$ and $\xi_N \propto \xi^{N-1}$.  The condition
$\beta=(3+n)/6$ can be understood further by examining individual
halos.  Using $c\propto M_*^{\beta_0}$, $M_*\propto a^{6/(n+3)}$, and
$\beta_0\approx \beta$ up to a logarithmic factor, we see that
$\beta=(3+n)/6$ gives approximately $c\propto a(t)$, which is
consistent with the simulation results in Bullock et al. (2000).  The
halo scale radius $R_s$ is then approximately constant in time in
physical coordinates, as expected for a stable system, and the
physical density $\rho$ of individual halos is indeed very close to
constant in time, more so for very small halos that form early ($\nu
\gg 1$, $ c \gg 1$) than for those that form recently ($\nu \approx
1$, $c\approx 1$).

Yano \& Gouda (1999) have found that for halo density $\rho\propto
r^{-\epsilon}$ with $\epsilon\le 3/2$ (such as Navarro et al. and
Moore et al.), the velocity parameter $v/Hr$ approaches 0 and the
two-point correlation function $\xi$ approaches a constant
(i.e. $\gamma=0$) in the nonlinear limit.  Neither behavior is seen in
our Figures because their derivation ignores the mass distribution
function $dn/dM$ and is therefore valid only for equal mass halos.

Our earlier paper (Ma \& Fry 2000a) has already shown signs of
departure from the stable clustering hypothesis in high resolution
$N$-body simulations.  In this paper, the specific values $\alpha=0$
and $\beta=(3+n)/6$ required for stable clustering provide additional
evidence that this long-cherished hypothesis may not be applicable in
all situations.  In such cases, one consequence is that the frequently
used fitting formulas for the nonlinear power spectrum (Hamilton et al
1991; Jain et al. 1995; Peacock \& Dodds 1996; Ma 1998) will need
modifications at high $k$, e.g., $k/k_{nl}\ga 50$ for the $n=-2$ model
and $k\ga 20\,h$ Mpc$^{-1}$ for cold dark matter models with a
cosmological constant (see Figs.~3 and 4 of Ma \& Fry 2000b).  We
believe the analytic halo model is a powerful tool that is providing
new insight into the nonlinear regime of gravitational clustering, the
most fundamental process in cosmology.

C.-P. M thanks Jim Peebles and Uros Seljak for useful discussions.
She acknowledges support of an Alfred P. Sloan Foundation Fellowship,
a Cottrell Scholars Award from the Research Corporation, a Penn
Research Foundation Award, and NSF grant AST 9973461.

\clearpage

\epsscale{.9} 
\plotone{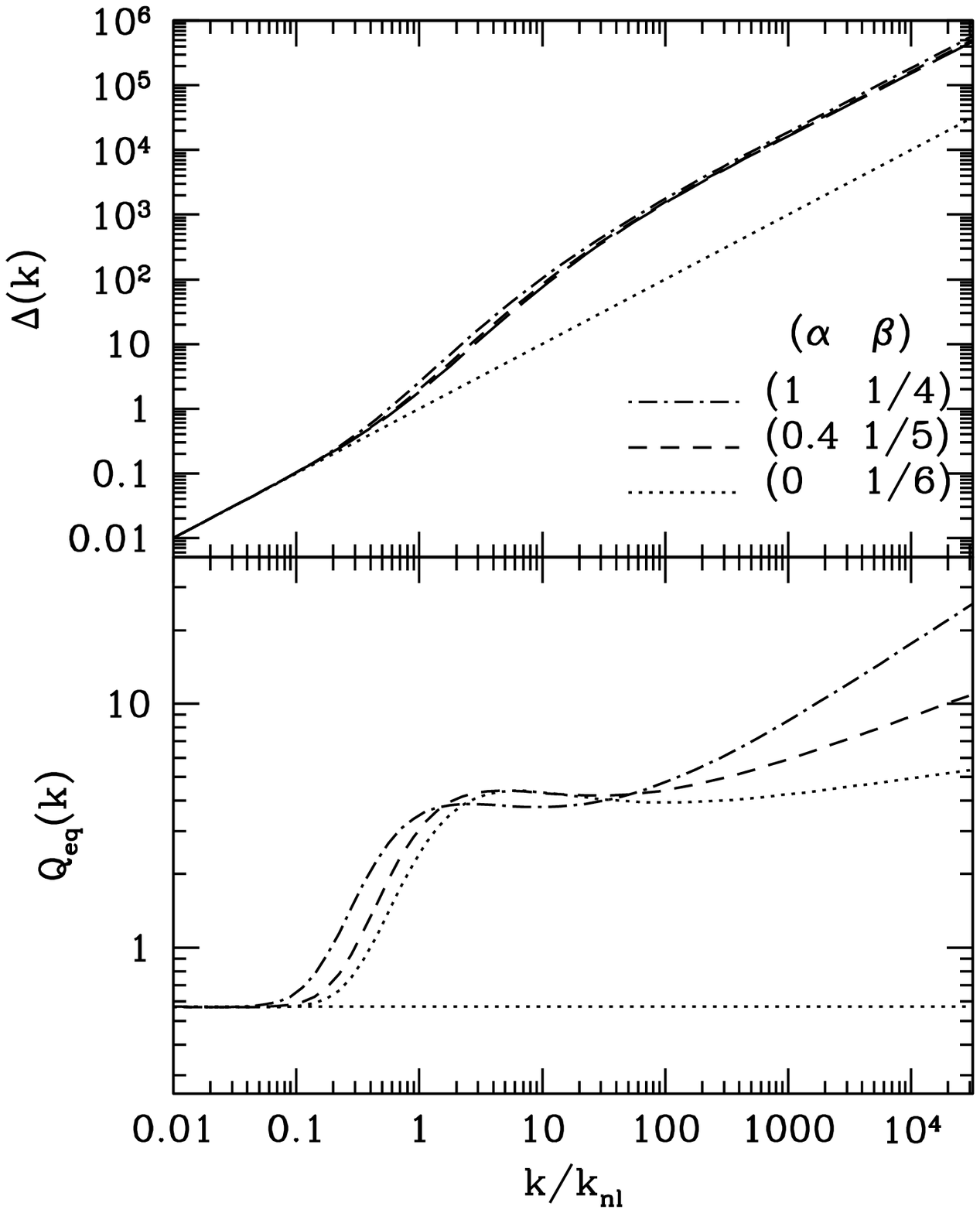} 
\figcaption {The power spectrum (upper panel) and three-point
amplitude (lower panel) given by the analytic halo model of Ma \& Fry
(2000b) for the $n=-2$ scale-free model.  Three sets of
($\alpha,\beta$) are shown, where $\alpha$ and $\beta$ parameterize
the halo mass function and halo concentration as defined in
eqs.~(\protect{\ref{alpha}}) and (\protect{\ref{beta}}).  (Note
$\beta_0$ in eq.~[\ref{beta}] is related to $\beta$ here by
$\beta\approx 0.8\beta_0$; see text.)  The Navarro et al. profile with
$c(M_*)=9$ is assumed.  Dotted straight lines show the lowest order
perturbative predictions.  While different combinations of
($\alpha,\beta$) give nearly identical $\Delta(k)$ which all agree
with the two-point stable clustering result (see eq.~[\ref{stable2}]),
the three-point $Q(k)$ has distinct high-$k$ behavior (see
eq.~[\ref{gamma3_model}]), and stable clustering holds to all orders
only if $\alpha=0$ and $\beta=(3+n)/6$.}

\epsscale{.95}  
\plotone{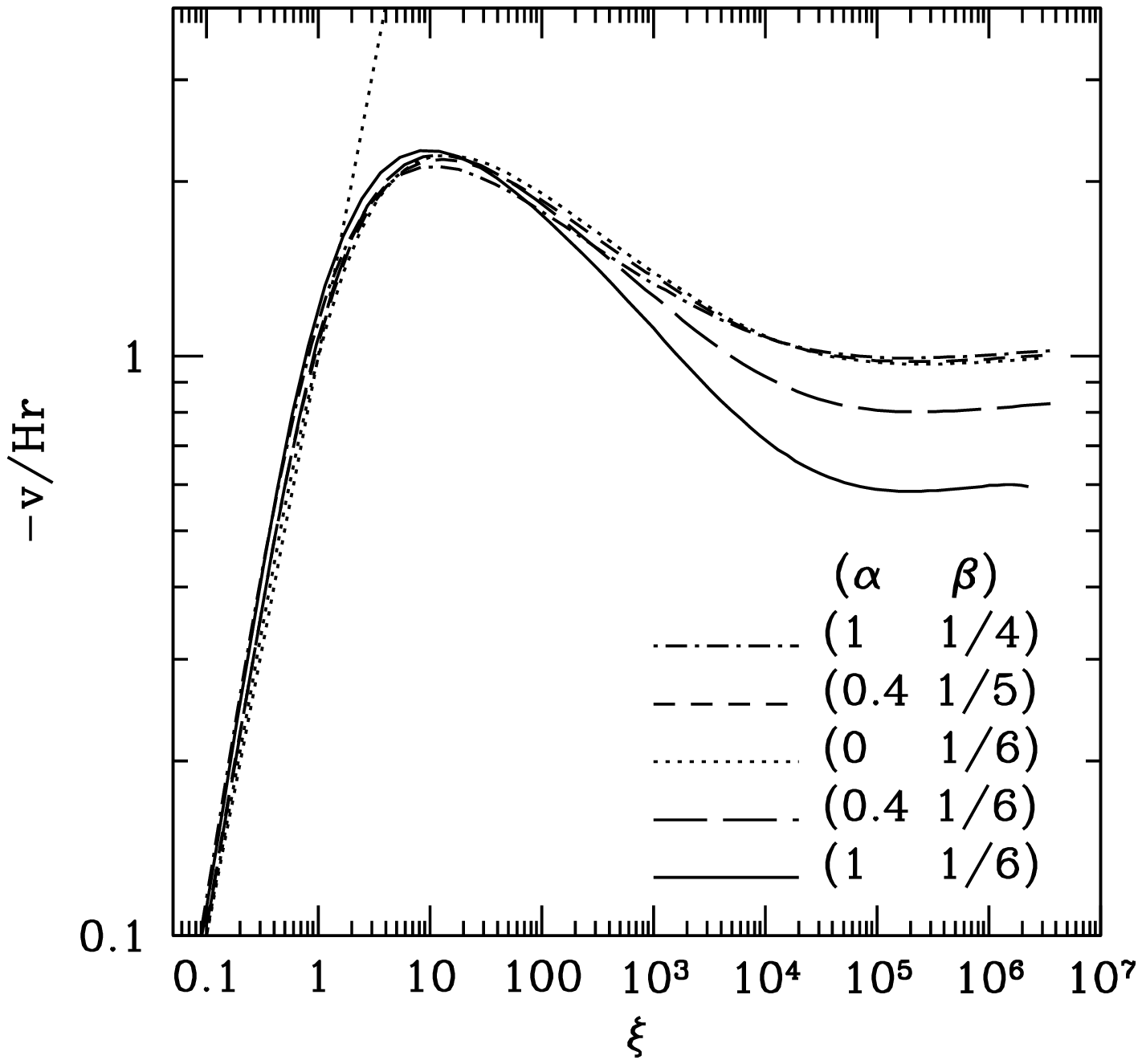} 
\figcaption {Prediction of the analytic halo model of Ma \& Fry (2000b)
for the ratio of pairwise peculiar velocity and Hubble flow, $-v/Hr$,
vs. the two-point correlation function $\xi$ for the $n=-2$ model.
The velocity ratio approaches a constant in the deeply nonlinear
regime (see eq.~[\ref{h_model}]), but the value is not always unity as
required by stable clustering.  For example, $-v/Hr\rightarrow 1$ at
$\xi\gg 1$ for $(\alpha, \beta)=(1,1/4), (0.4,1/5), (0,1/6)$, but it
reaches a smaller value for $\alpha=0.4$ or 1 and $\beta=1/6$, which
are the range of values found in simulations.  Dotted straight line
shows the linear theory prediction, which is followed accurately at
$\xi\la 1$.}

\end{document}